\documentclass[aps,prd,groupedaddress,preprint,nofootinbib,showpacs]{revtex4}
\usepackage{graphicx,epsf,amssymb,amsbsy,amsfonts,amssymb,amsmath}
\usepackage[dvips]{color}

\def\eqn#1{eq.~(\ref{#1})}

\def\eqns#1#2{eqs.~(\ref{#1}) and~(\ref{#2})}

\def\fig#1{fig.~{\ref{#1}}}

\begin{document}

\hfill SLAC--PUB--13498

\title{Resonance--Continuum Interference 
in Light Higgs Boson Production at a Photon Collider}

\vspace{2cm}

\author{Lance J. Dixon and Yorgos Sofianatos}

\affiliation{
SLAC National Accelerator Laboratory \\
Stanford University, Stanford, CA 94309, USA}

\date{December, 2008}

\begin{abstract}
We study the effect of interference between the Standard Model Higgs boson
resonance and the continuum background in the
process $\gamma\gamma\rightarrow H\rightarrow b\bar{b}$ at a photon
collider.  Taking into account virtual gluon exchange between
the final-state quarks, we calculate the leading corrections to the
height of the resonance for the case of a light ($m_H<160$ GeV)
Higgs boson. We find that the interference is destructive and around
0.1--0.2\% of the peak height, depending on the mass of the Higgs 
and the scattering angle.  This suppression is smaller by an order of 
magnitude than the anticipated experimental accuracy at a photon collider.  
However, the fractional suppression can be significantly larger if
the Higgs coupling to $b$ quarks is increased by physics beyond
the Standard Model.
\end{abstract}


\maketitle

%%%%%%%%%%%%%%%%%%%%%%%%%%%%%%%%

The Standard Model of Particle Physics (SM) has been very successful
in describing a wide range of elementary particles phenomena to high
accuracy. A key ingredient of the model is the scalar Higgs field,
responsible for electroweak symmetry breaking and for generating the
masses of essentially all massive elementary 
particles~\cite{Higgs:1966ev,Englert:1964et,Guralnik:1964eu}. 
Similar fields exist in extensions of the SM, such as the Minimal Supersymmetric
Standard Model (MSSM). In the SM, the Higgs boson is the only
particle that remains undiscovered, and its properties are
determined by its mass.  It is a main goal of current and future high
energy physics experiments to identify the Higgs boson and explore the
details of the Higgs sector. In particular, the discovery of the Higgs
boson could take place at Run II of the Tevatron at Fermilab; if not
there, then at the Large Hadron Collider (LHC) at CERN.  Precise measurements
of its properties will be one of the tasks of the proposed
International Linear Collider (ILC).  There is an option to use
the ILC as a photon collider, by backscattering laser light off of
the high energy electron beams. The high energy, highly polarized
photons produced in this way can be used to study the various Higgs 
couplings to very high 
accuracy~\cite{Djouadi:2007ik,Badelek:2001xb,Bechtel:2006mr,%
Krawczyk:2003hb,DeRoeck:2003gv,Telnov:1995hc}.

The mass of the Higgs boson in the SM and MSSM has already been constrained by
experiment to a range well within the reach of the aforementioned
designed machines. Precision electroweak measurements have put an
upper bound on the allowed values for its mass, $m_H\lesssim 170$
GeV at 95\% confidence level in the
SM~\cite{Renton:2008ua,Alcaraz:2007ri}.  In the MSSM the Higgs boson
mass obeys the bound $m_H\leq m_Z$ at tree level; radiative
corrections increase this limit to about 
135~GeV~\cite{Carena:2000dp,Espinosa:2000df,Brignole:2001jy}.  The mass of
the Higgs boson has also been bounded from below via the Higgs-strahlung
process $e^{+}e^{-}\rightarrow HZ$ at LEP2, with $m_H\gtrsim 114.1$
GeV in the SM and $m_H\gtrsim 91.0$ GeV in the 
MSSM~\cite{Barate:2000ts,Acciarri:2000hv,Abbiendi:2000ac,Abreu:2000fw,:2001xwa,:2001xx}.

At a photon collider, among the two possible modes, $\gamma\gamma$
and $e\gamma$, the former is especially useful for Higgs physics.
For $m_H<140$ GeV, the most important channel involves Higgs
production via photon fusion, $\gamma\gamma\rightarrow H$, followed
by the decay $H\rightarrow b\bar{b}$~\cite{Gunion:1992ce,Borden:1993cw}. 
The advantage of this channel is
that the amplitude for the continuum $\gamma\gamma\rightarrow
b\bar{b}$ background to the Higgs signal is suppressed by a factor
of $\mathcal{O}\left(m_b/\sqrt{s_{\gamma\gamma}}\right)$ when the
initial-state photons are in a $J_z=0$ state.  The production
of a light SM Higgs boson through this process has been studied in a
series of papers, including the radiative QCD corrections to the
signal and to the 
backgrounds~\cite{Djouadi:1990aj,Melnikov:1993tj,Borden:1994fv,%
Jikia:1994vt,Kamal:1995ct,Khoze:1995mn,Fadin:1997sn,%
Melles:1998gu,Melles:1998xd,Melles:1998rp,Melles:1999xd,%
SoldnerRembold:2000pb,Akhoury:2002ag,Khoze:2006um}. 
The anticipated experimental uncertainty in the measurement of the 
partial Higgs width, 
$\Gamma(H\rightarrow\gamma\gamma)\times\mathrm{Br}(H\rightarrow b\bar{b})$, 
assuming an integrated luminosity of 80 fb$^{-1}$ in the
high energy peak, is about 2\% for $m_H<140$ GeV~\cite{
Monig:2007py,Niezurawski:2005cp,Niezurawski:2005xw,Niezurawski:2003iu,%
Niezurawski:2002aq,Ginzburg:2001wj,Ginzburg:2001ss,Melles:1999xd,%
SoldnerRembold:2000pb,Bechtel:2006mr}.

It is important to know that no other effect can contaminate the
$b\bar{b}$ signal at the 1\% level. A possible concern studied in
this paper is the interference between the resonant Higgs amplitude
$\gamma\gamma\rightarrow H\rightarrow b\bar{b}$, and the continuum
$\gamma\gamma\rightarrow b\bar{b}$ process. Similar effects have
been studied previously in $gg\rightarrow H\rightarrow t\bar{t}$ at
a hadron collider~\cite{Dicus:1994bm}, and in
$\gamma\gamma\rightarrow H\rightarrow W^{+}W^{-}$, $ZZ$ and $t\bar{t}$ at
a photon collider~\cite{Morris:1993bx,Niezurawski:2002jx,Asakawa:1999gz}.
These studies assumed a Higgs boson sufficiently heavy that its
width was at the GeV scale due to on-shell decays to 
$W^{+}W^{-}$, $ZZ$ and $t\bar{t}$. 
In the MSSM, interference effects in 
$\gamma\gamma\rightarrow H\rightarrow b\bar{b}$, as well as
in decays to several other final states, were taken into account,
including also Sudakov 
resummation~\cite{Muhlleitner:2000jj,Muhlleitner:2001kw}.
However, explicit results separating out the interference 
contributions in the SM were not presented.
The significance of such interference effects in CP asymmetries 
for various channels of MSSM Higgs production and decay at a 
photon collider has also been explored~\cite{Ellis:2004hw}.
In the case of a light SM Higgs boson, with an MeV-scale width, 
the interference in $gg\rightarrow H\rightarrow\gamma\gamma$ 
was considered at the 
LHC~\cite{Dixon:2003yb}. Resonance-continuum interference effects are
usually negligible for a narrow resonance, and for $m_H<$ 150 GeV
the width $\Gamma_H$ is less than 17 MeV in the SM.\footnote{%
In the MSSM, the widths of light Higgs bosons may be GeV-scale
if $\tan\beta$ is large, {\it e.g.} as considered in 
ref.~\cite{Ellis:2004hw}.}
However, the $\gamma\gamma\rightarrow H\rightarrow b\bar{b}$ resonance is also
rather weak, since it consists of a one-loop production amplitude.
Therefore a tree-level, or even one-loop, continuum amplitude can
potentially compete with it, especially since the tree-level
$\mathcal{O}\left(m_b/\sqrt{s_{\gamma\gamma}}\right)$ suppression of
the $\gamma\gamma\rightarrow b\bar{b}$ continuum amplitude is absent
at one loop. In the analogous case of
$gg\rightarrow H \rightarrow\gamma\gamma$, 
a suppression of $\sim$ 5\% was found due to 
continuum interference~\cite{Dixon:2003yb}.

In the SM, the production amplitude $\gamma\gamma\rightarrow H$
proceeds at one loop and is dominated by a $W$ boson in the loop,
with some top quark contribution as well. The decay $H\rightarrow
b\bar{b}$ and the continuum $\gamma\gamma\rightarrow b\bar{b}$
amplitudes proceed at tree level. For $m_H<160$ GeV, the Higgs boson is
below the $t\bar{t}$ and $WW$ thresholds, so the resonant amplitude
is predominantly real ({\it i.e.}, has no absorptive part), apart from the
relativistic Breit-Wigner factor. The full $\gamma\gamma\rightarrow
b\bar{b}$ amplitude is a sum of resonance and continuum terms,
\begin{equation}
\mathcal{A}_{\mathrm{total}} = \frac{ -
\mathcal{A}_{\gamma\gamma\rightarrow H}\mathcal{A}_{H\rightarrow
b\bar{b}}}{s - m_H^2 + im_H\Gamma_H} +
\mathcal{A}_{\gamma\gamma\rightarrow b\bar{b}} \,,
\label{full amplitude}
\end{equation}
where $s = s_{\gamma\gamma}$ is the photon-photon invariant mass.
The interference term in the cross section is given by
\begin{equation}
\begin{split}
\delta\sigma_{\gamma\gamma\rightarrow H\rightarrow b\bar{b}} = &-
2(s - m_H^2)
\frac{\mathrm{Re} \left\{
\mathcal{A}^*_{\gamma\gamma\rightarrow H}
\mathcal{A}^*_{H\rightarrow b\bar{b}}
\mathcal{A}_{\gamma\gamma\rightarrow b\bar{b}}\right\}}
{(s - m_H^2)^2 + m_H^2\Gamma_H^2} \\ 
& + 2 m_H\Gamma_H
\frac{\mathrm{Im}\left\{
\mathcal{A}^*_{\gamma\gamma\rightarrow H}
\mathcal{A}^*_{H\rightarrow b\bar{b}}
\mathcal{A}_{\gamma\gamma\rightarrow b\bar{b}} \right\}}
{(s - m_H^2)^2 + m_H^2\Gamma_H^2} \,. 
\label{delta sigma}
\end{split}
\end{equation}

Since the intrinsic Higgs width $\Gamma_H$ is much narrower than the
spread of the luminosity spectrum in $\sqrt{s}$~\cite{Telnov:1995hc} and the
experimental resolution $\delta m_H\sim$ 0.5 GeV~\cite{DeRoeck:2003gv}, 
the observable interference effect is the integral
over $s$ across the entire linewidth. Neglecting the tiny 
$s$-dependence of 
$\mathrm{Re}\left\{\mathcal{A}^*_{\gamma\gamma\rightarrow H}
\mathcal{A}^*_{H\rightarrow b\bar{b}}
\mathcal{A}_{\gamma\gamma\rightarrow b\bar{b}}\right\}$,
the integral of the first ``real'' term vanishes, as it is an odd
function of $s$ around $m_H^2$. The second ``imaginary'' term is an
even function of $s$ around $m_H^2$ and therefore survives the
integration. However, it requires a relative phase between the
resonant and continuum amplitudes. As described above, in the SM
the resonant amplitude is mainly real, apart from the Breit-Wigner
factor.  The tree level continuum $\gamma\gamma\rightarrow
b\bar{b}$ amplitude is also real.
\begin{figure}[t]
\begin{center}
\scalebox{0.7}{
\includegraphics{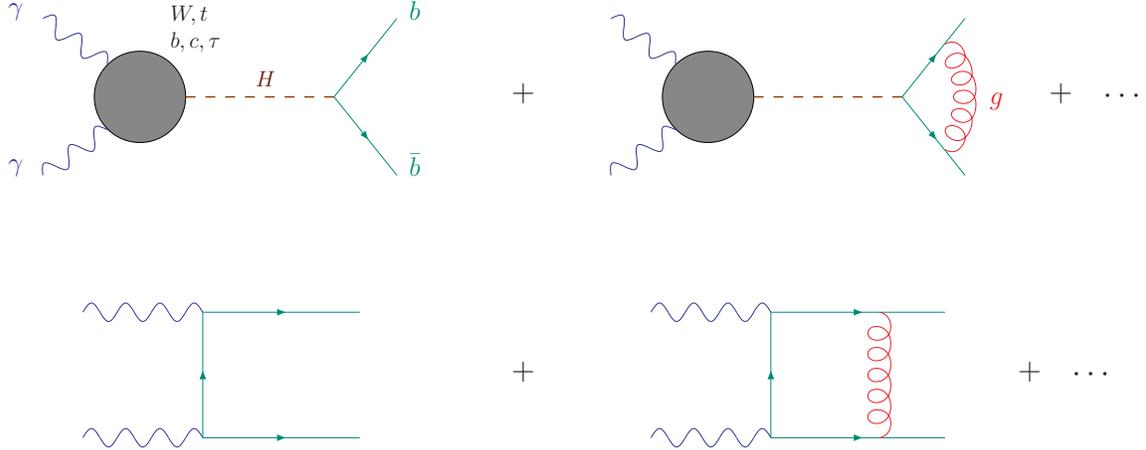}}
\caption{{\small Feynman diagrams contributing to the interference
of $\gamma\gamma\rightarrow H\rightarrow b\bar{b}$ (upper row) with the
continuum background (lower row) up to order $\mathcal{O}\left(\alpha_s\right)$.
Only one diagram is shown at each loop order, for each amplitude.
The blob contains $W$ and $t$ loops, and small contributions from
lighter charged fermions.}} \label{interference}
\end{center}
\end{figure}
The imaginary parts  of the $H\rightarrow b\bar{b}$ and
$\gamma\gamma\rightarrow b\bar{b}$ amplitudes arise at one loop,
when we include the exchange of a gluon between the $b$ and
$\bar{b}$ quarks. These contributions are shown schematically in
\fig{interference}. In fact, each amplitude individually has an infrared
divergence from the soft-gluon exchange that builds up the Coulomb phase.
However, the divergence cancels in the relative phase entering
$\mathrm{Im}\left\{\mathcal{A}^*_{\gamma\gamma\rightarrow H}
\mathcal{A}^*_{H\rightarrow b\bar{b}}
\mathcal{A}_{\gamma\gamma\rightarrow b\bar{b}}\right\}$.
Thus we are left with a finite contribution to
$\delta\sigma_{\gamma\gamma\rightarrow H\rightarrow b\bar{b}}$ in
\eqn{delta sigma}.  To compute the fractional interference
correction to the resonance, we divide \eqn{delta sigma}
for $\delta\sigma_{\gamma\gamma\rightarrow H\rightarrow b\bar{b}}$
by the square of the resonant amplitude in \eqn{full amplitude}. 
We then expand all the amplitudes in $\alpha_s$, obtaining
\begin{equation}
\delta\equiv 
\frac{\delta\sigma_{\gamma\gamma\rightarrow H\rightarrow b\bar{b}}}
{\sigma_{\gamma\gamma\rightarrow H\rightarrow b\bar{b}}}
= 2m_H\Gamma_H \, \mathrm{Im}\left\{
\frac{\mathcal{A}^\mathrm{tree}_{\gamma\gamma\rightarrow b\bar{b}}}
{\mathcal{A}^{\left(1\right)}_{\gamma\gamma\rightarrow H}
\mathcal{A}^\mathrm{tree}_{H\rightarrow b\bar{b}}}
\ \left[1 +
\frac{\mathcal{A}^{\left(1\right)}_{\gamma\gamma\rightarrow b\bar{b}}}
{\mathcal{A}^\mathrm{tree}_{\gamma\gamma\rightarrow b\bar{b}}} 
- \frac{\mathcal{A}^{(2)}_{\gamma\gamma\rightarrow H}}
{\mathcal{A}^{\left(1\right)}_{\gamma\gamma\rightarrow H}}
- \frac{\mathcal{A}^{\left(1\right)}_{H\rightarrow b\bar{b}}}
{\mathcal{A}^\mathrm{tree}_{H\rightarrow b\bar{b}}} \right]\right\},
\end{equation}
where the superscript $(l)$ denotes the number of loops
($l=1,2$) for each term in the expansion, {\it e.g.}
$\mathcal{A}_{\gamma\gamma\rightarrow H} =
\mathcal{A}^{\left(1\right)}_{\gamma\gamma\rightarrow H} +
\mathcal{A}^{\left(2\right)}_{\gamma\gamma\rightarrow H} +
\ldots$.

Taking into account that the tree amplitude
$\mathcal{A}^\mathrm{tree}_{H\rightarrow b\bar{b}}$ has no
absorptive part, we can rewrite $\delta$ as
\begin{equation}
\begin{split}
\delta = \frac{2m_H\Gamma_H}
{\left|\mathcal{A}^\mathrm{tree}_{H\rightarrow b\bar{b}}\right|^2}
\ \mathrm{Im}&\left\{ \frac{1}
{\mathcal{A}^{\left(1\right)}_{\gamma\gamma\rightarrow H}}
\Bigg[\mathcal{A}^\mathrm{tree}_{\gamma\gamma\rightarrow
b\bar{b}}\mathcal{A}^\mathrm{{}^*tree}_{H\rightarrow b\bar{b}} +
\mathcal{A}^\mathrm{{}^*tree}_{H\rightarrow
b\bar{b}}\mathcal{A}^{\left(1\right)}_{\gamma\gamma\rightarrow
b\bar{b}}\right.\\ &\left. \null -
\mathcal{A}^\mathrm{tree}_{\gamma\gamma\rightarrow
b\bar{b}}\mathcal{A}^\mathrm{{}^*tree}_{H\rightarrow b\bar{b}}
\frac{\mathcal{A}^{(2)}_{\gamma\gamma\rightarrow H}}
{\mathcal{A}^{\left(1\right)}_{\gamma\gamma\rightarrow H}} -
\mathcal{A}^\mathrm{tree}_{\gamma\gamma\rightarrow
b\bar{b}}\mathcal{A}^{{}^*\left(1\right)}_{H\rightarrow
b\bar{b}}\Bigg]\right\}.
\end{split}
\end{equation}
We neglect the two-loop amplitude 
$\mathcal{A}^{(2)}_{\gamma\gamma\rightarrow H}$,
because in the SM it is dominantly real for $m_H<2m_W$, 
like $\mathcal{A}^{(1)}_{\gamma\gamma\rightarrow H}$, 
up to small contributions from loops of lighter fermions.
We also separate out the contribution from the small imaginary part of
$\mathcal{A}^{\left(1\right)}_{\gamma\gamma\rightarrow H}$, obtaining
the expression
\begin{equation}
\begin{split}
\delta = \frac{2m_H\Gamma_H}
{\left| \mathcal{A}^\mathrm{tree}_{H\rightarrow b\bar{b}} \right|^2}
\Bigg[ &- 
\frac{\mathcal{A}^\mathrm{tree}_{\gamma\gamma\rightarrow b\bar{b}}
\mathcal{A}^\mathrm{^{*}tree}_{H\rightarrow b\bar{b}}}
{\left|\mathcal{A}^{\left(1\right)}_{\gamma\gamma\rightarrow H}\right|^2}
\ \mathrm{Im}\left\{
\mathcal{A}^{\left(1\right)}_{\gamma\gamma\rightarrow H}\right\} \\
&+ \frac{1}
{\mathrm{Re}\left\{
\mathcal{A}^{\left(1\right)}_{\gamma\gamma\rightarrow H}\right\}}
\ \mathrm{Im}\left\{
\mathcal{A}^\mathrm{^{*}tree}_{H\rightarrow b\bar{b}}
\mathcal{A}^{\left(1\right)}_{\gamma\gamma\rightarrow b\bar{b}}
- \mathcal{A}^\mathrm{tree}_{\gamma\gamma\rightarrow b\bar{b}}
\mathcal{A}^{^{*}\left(1\right)}_{H\rightarrow b\bar{b}}
\right\}\Bigg]. \label{delta simplified}
\end{split}
\end{equation}
The two photons in \eqn{delta simplified} are taken to have 
identical helicity in all amplitudes, so that $J_z=0$ as required for 
interference with the production of the scalar Higgs boson.

\begin{figure}[t]
\begin{center}
\scalebox{0.5}{
\includegraphics{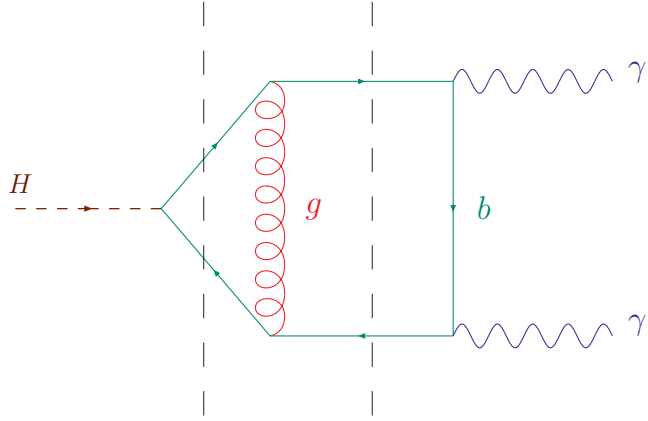}}
\caption{{\small Feynman diagram for the calculation of the
interference of $\gamma\gamma\rightarrow H\rightarrow b\bar{b}$ with
the continuum background up to order
$\mathcal{O}\left(\alpha_s\right)$. The unitarity cuts indicated by
dashed vertical lines are used to compute the imaginary parts of the
various amplitudes.}} \label{diagram}
\end{center}
\end{figure}

We determine the imaginary parts of the two terms in the braces in
the second line of \eqn{delta simplified} by analyzing
the unitarity cuts of the diagram in~\fig{diagram}. 
The first term, 
$\mathrm{Im}\Bigl\{
\mathcal{A}^\mathrm{^{*}tree}_{H\rightarrow b\bar{b}}
\mathcal{A}^{\left(1\right)}_{\gamma\gamma\rightarrow b\bar{b}}\Bigr\}$,
comes from interpreting the $b$ quarks crossing the left cut as
the actual final-state $b$ quarks, emerging at a fixed scattering angle $\theta$.
The imaginary part of the tensor box integral to the right of the left cut 
is associated with $b$-quark rescattering; thus one integrates over the 
$b$ momenta crossing the right cut.   The second term,
$- \mathrm{Im}\Bigl\{ 
\mathcal{A}^\mathrm{tree}_{\gamma\gamma\rightarrow b\bar{b}}
\mathcal{A}^{^{*}\left(1\right)}_{H\rightarrow b\bar{b}} \Bigr\}$,
comes from exchanging the roles of the left and right cuts.

We use \texttt{FORM}~\cite{Vermaseren:2000nd} for symbolic manipulations,
and the decomposition of the scalar box integral into a
six-dimensional scalar box plus scalar triangle
integrals~\cite{Bern:1992em}.  In the expressions below, we use the same
notation as in ref.~\cite{Bern:1992em}; primed quantities
correspond to particular tensor integrals. After cancelling the
divergent parts arising from both terms (associated with the scalar
triangle integral $I_3^{(2)}[1]$), the finite imaginary parts
are given by
\begin{eqnarray}
\mathrm{Im}\left\{
\mathcal{A}^\mathrm{^{*}tree}_{H\rightarrow b\bar{b}}
\mathcal{A}^{(1)\,\mathrm{fin}}_{\gamma\gamma\rightarrow b\bar{b}}\right\}
&=& \frac{8 Q_b^2 \alpha \alpha_s m_b}{m_H^2 v}
\Bigg[ 2 m_b m_H^2 \left(m_H^4 - 6 m_b^2 m_H^2 + 8m_b^4\right)
\mathrm{Im}\left\{I_4\left[1\right]\right\} \nonumber\\
&&\hskip0.5cm \null
 - 8 m_b^3 m_H^2 \,\mathrm{Im}\left\{I_3^{\left(4\right)}\left[1\right]\right\}
 - 4 m_b \left(m_H^2 - 4m_b^2\right)
\mathrm{Im}\left\{I_3^{\left(2\right)\prime}\right\}\Bigg]\nonumber\\
&& \hskip0.5cm \null
+\ \big( \! \cos\theta\rightarrow - \cos\theta\big),
\label{box side}\\
\mathrm{Im}\left\{
\mathcal{A}^\mathrm{tree}_{\gamma\gamma\rightarrow b\bar{b}}
\mathcal{A}^{^{*}(1)\,\mathrm{fin}}_{H\rightarrow b\bar{b}}\right\} 
&=& \frac{8 Q_b^2 \alpha \alpha_s m_b}{m_H^2 v}
\Bigg[
- 4 m_b m_H^2 \mathrm{Im}\left\{I_3^{\left(2\right)\prime}\right\}
\nonumber\\
&&\hskip0.5cm \null
 + \frac{4m_b}{t - m_b^2}\left[ 2 t^2 + (m_H^2 - 4 m_b^2) t
 + m_b^2 m_H^2 + 2 m_b^4 \right]
\mathrm{Im}\left\{I_2^{\left(2,4\right)}\left[1\right]\right\}\Bigg] 
\nonumber\\
&& \hskip0.5cm \null
+\ \big( \! \cos\theta\rightarrow - \cos\theta\big),
\label{triangle side}
\end{eqnarray}
where
\begin{eqnarray}
\mathrm{Im}\left\{I_4\left[1\right]\right\} &=& 
\frac{1}{2} \left[
c_4\,\mathrm{Im}\left\{I_3^{\left(4\right)}\left[1\right]\right\} 
- c_0\,\mathrm{Im}\left\{I_4^{D=6-2\epsilon}\left[1\right]\right\}\right], 
\\
\mathrm{Im}\left\{I_3^{\left(4\right)}\left[1\right]\right\} &=&
\frac{\pi}{m_H^2} \ln\left(\frac{1 + \beta}{1 - \beta} \right),
\end{eqnarray}
\begin{eqnarray}
\mathrm{Im}\left\{I_3^{\left(2\right)\prime}\right\} &=&
\pi\left[\beta + \frac{2\left(t +
m_b^2\right)}{\beta m_H^2}\right], \\
\mathrm{Im}\left\{I_2^{\left(2,4\right)}\left[1\right]\right\} &=&
\pi\beta \,, %\\
\end{eqnarray}
and
\begin{eqnarray}
\mathrm{Im}\left\{I_4^{D=6-2\epsilon}\left[1\right]\right\} &=&
\pi \Biggl[
\frac{\left(1 + \beta\right)\ln\Bigl[ 
\frac{m_H^2\left(1 + \beta\right)} {2\left(m_b^2 - t\right)} \Bigr]}
{m_H^2\left(1 + \beta\right) + 2\left(t - m_b^2\right)}
 - \frac{\left(1 - \beta\right) \ln\Bigl[ 
\frac{m_H^2\left(1 - \beta\right)}{2\left(m_b^2 - t\right)} \Bigr]}
{m_H^2\left(1 - \beta\right) + 2\left(t - m_b^2\right)} \Biggr], \\
c_4 &=& \,\frac{2m_H^2\left(t + m_b^2\right)}
{\left(t - m_b^2\right)^2\left(4m_b^2 -m_H^2\right)}\,, \\
c_0 &=& 4\,\frac{t^2 + t\left(m_H^2 - 2m_b^2\right) + m_b^4}
{\left(t - m_b^2\right)^2\left(4m_b^2 - m_H^2\right)} \,.
\end{eqnarray}
In the expressions above,
\begin{equation}
\beta\equiv\sqrt{1 - \frac{4m_b^2}{m_H^2}}\,,
\end{equation}
and
\begin{equation}
t = m_b^2 - \frac{m_H^2}{2}\left(1+\beta\cos\theta\right),
\end{equation}
where $\theta$ is the $\gamma\gamma\rightarrow b\bar{b}$
center-of-mass scattering angle.
The terms in \eqns{box side}{triangle side} that are obtained by
substituting $\cos\theta\rightarrow - \cos\theta$
(or, equivalently, $t\rightarrow 2m_b^2 - m_H^2 - t$)  
arise from a diagram like that in~\fig{diagram}, 
but with the two photons exchanged.

It is worth noting that the absence of bubble integrals from
\eqn{box side} is due to a cancellation among the scalar and tensor
bubble terms, and that the tensor triangle contribution in
\eqn{triangle side} has been expressed in terms of the tensor
triangle integral $I_3^{\left(2\right)\prime}$ appearing in \eqn{box side}.
After adding the terms with $\cos\theta\rightarrow - \cos\theta$,
the contributions from $I_3^{\left(2\right)\prime}$ drop out.
Simplifying, we get
\begin{eqnarray}
&&\mathrm{Im}\left\{
\mathcal{A}^\mathrm{^{*}tree}_{H\rightarrow b\bar{b}}
\mathcal{A}^{\left(1\right)}_{\gamma\gamma\rightarrow b\bar{b}}
- \mathcal{A}^\mathrm{tree}_{\gamma\gamma\rightarrow b\bar{b}}
\mathcal{A}^{^{*}\left(1\right)}_{H\rightarrow b\bar{b}}\right\} 
= 32 \pi Q_b^2 \alpha \alpha_s \frac{m_b^2}{v} \Bigg\{
\nonumber\\&&\hskip1cm 
(m_H^2 - 2m_b^2)\left[1 + \frac{m_H^2t}{(m_b^2 - t)^2}\right] 
\left[
\frac{(1 + \beta)\ln\left[\frac{m_H^2(1 + \beta)}{2(m_b^2 - t)}\right]}
{m_H^2(1 + \beta) + 2(t - m_b^2)}
- \frac{(1 - \beta)\ln\left[\frac{m_H^2(1 - \beta)}{2(m_b^2 - t)}\right]}
{m_H^2(1 - \beta) + 2(t - m_b^2)}\right]
\nonumber\\ &&\hskip1cm
\null - \left[\frac{(m_H^2 - 2m_b^2)(t + m_b^2)}{2(m_b^2 - t)^2} 
+ \frac{2m_b^2}{m_H^2} \right]
\ln\left(\frac{1 + \beta}{1 - \beta}\right)
+ \frac{2 \beta m_b^2}{m_b^2 - t}\Bigg\}
\nonumber\\ &&\hskip1cm
\null + \big( \! \cos\theta\rightarrow - \cos\theta\big) \,.
\label{ImExactMass}
\end{eqnarray}
To evaluate \eqn{delta simplified}, we also need the one-loop amplitude
for $H\rightarrow\gamma\gamma$~\cite{Ellis:1975ap,Resnick:1973vg},
\begin{equation}
\mathcal{A}^{\left(1\right)}_{\gamma\gamma\rightarrow H} =
\frac{\alpha m_H^2}{4\pi v} \Bigg[3\sum_{q=t,b,c}Q_q^2A_q^H\left(
\frac{4m_q^2}{m_H^2} \right) + A_q^H\left( \frac{4m_\tau^2}{m_H^2}
\right) + A_W^H\left( \frac{4m_W^2}{m_H^2} \right)\Bigg],
\label{Hgammagamma}
\end{equation}
with
\begin{align}
A_q^H\left(x\right) &= 2x\left[1 + \left( 1 -
x\right)f\left(x\right)\right], \\
A_W^H\left(x\right) &= - x\left[3 + \frac{2}{x} +
3\left(2 - x\right)f\left(x\right)\right], \\
f\left(x\right) &= \left\{\begin{array}{ll}\arcsin^2\left(
\frac{1}{\sqrt{x}} \right),&x\geqslant 1\, , \\ - \frac{1}{4}
\left[\ln\left( \frac{1 + \sqrt{1-x}}{1 - \sqrt{1-x}} \right) -
i\pi\right]^2,&x<1\, ,\end{array}\right.
\end{align}
and the tree amplitudes~\cite{Gunion:1992ce}
\begin{align}
\mathcal{A}^\mathrm{tree}_{H\rightarrow b\bar{b}} &=
\sqrt{6}\frac{m_b}{v}\sqrt{m_H^2-4m_b^2}\, ,\label{tree continuum}\\
\mathcal{A}^\mathrm{tree}_{\gamma\gamma\rightarrow b\bar{b}} &=
8\sqrt{6}\pi\alpha Q_b^2
\frac{\sqrt{1-\beta^4}}{1-\beta^2\cos^2\theta}\,.\label{tree Hbb}
\end{align}
Here we note that the color factors have been included in 
eqns.~\eqref{box side}, \eqref{triangle side}, \eqref{tree continuum} and
\eqref{tree Hbb};  the respective ``amplitudes'' are really the
square roots of cross sections, summed over the $b$ quark colors
and spins, for identical-helicity photons.

In the limit of small $m_b$, we can expand the contribution to
$\delta$ coming from the
$\mathcal{A}^{\left(1\right)}_{\gamma\gamma\rightarrow b\bar{b}}$
and $\mathcal{A}^{\left(1\right)}_{H\rightarrow b\bar{b}}$ phases
around $m_b = 0$. This approximation is excellent for almost
all scattering angles, because $m_b\ll\sqrt{s_{\gamma\gamma}}$.
We obtain the following formula,
\begin{equation}
\delta \approx 
\frac{128\pi Q_b^2 \alpha \alpha_s m_H \Gamma_H }{v} 
\, m_b^2 \, 
\frac{ 2\ln\bigl(\frac{m_H}{2m_b}\bigr)
+ 2\ln\left(\sin\theta\right) 
+ \ln\left( \frac{1 - \cos\theta}{1 + \cos\theta} \right)\cos\theta}
{\sin^2\theta\ \left|\mathcal{A}^\mathrm{tree}_{H\rightarrow b\bar{b}}\right|^2
\mathrm{Re} \bigl\{\mathcal{A}^{\left(1\right)}_{\gamma\gamma\rightarrow H}\bigr\}}
\ +\ \mathcal{O}\left(m_b^4\right).
\label{delta approx}
\end{equation}

\begin{figure}[t]
\begin{center}
\rotatebox{270}{\scalebox{0.63}{
\includegraphics{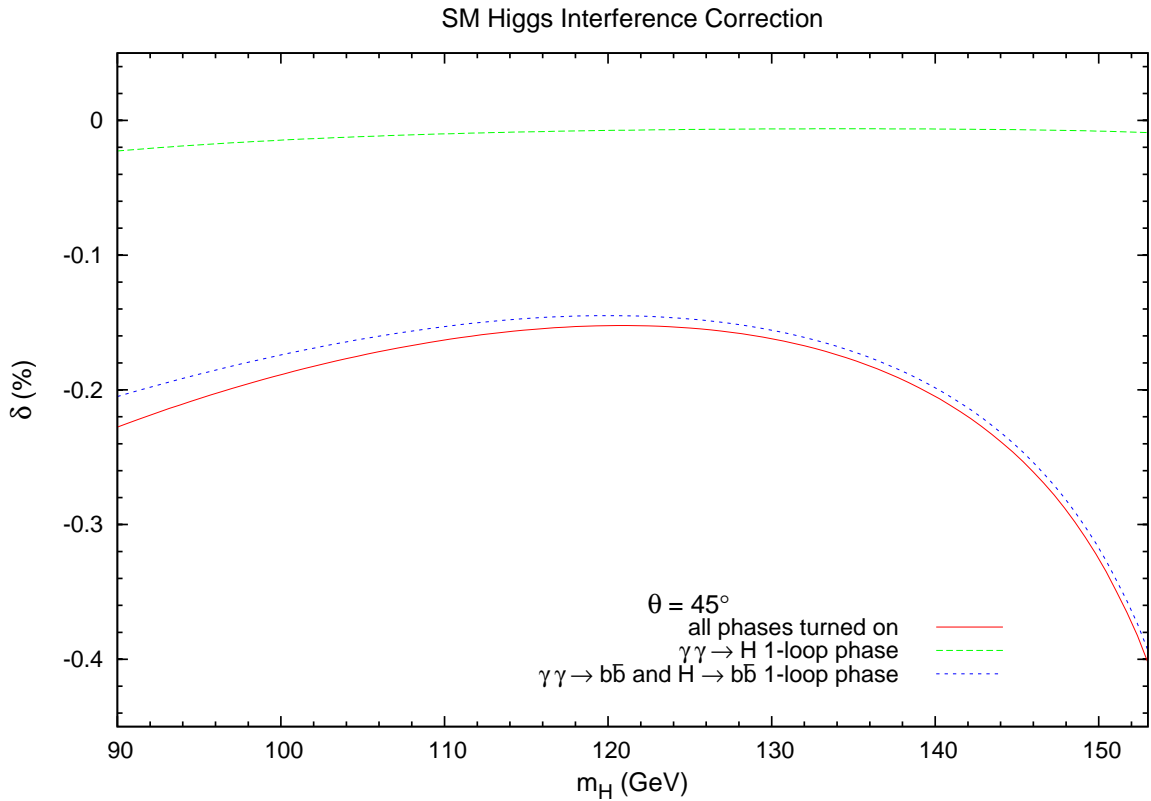}}}
\caption{{\small The percentage reduction of the SM Higgs signal as
a function of the Higgs boson mass, for center-of-mass scattering angle
$\theta = 45^{\circ}$. The solid curve represents the result with
all phases turned on; the dashed curves turn on different component
phases each time. The effect is stronger for
a higher mass Higgs boson.}} \label{delta vs mH}
\end{center}
\end{figure}

We evaluate $\delta$ by letting $\alpha = 1/137.036$, $\alpha_s =
0.119$, $v=246$~GeV, $m_t = 171.2$~GeV, $m_b = 4.24$ GeV, $m_c =
1.2$ GeV, $m_{\tau} = 1.78$ GeV, and $m_W = 80.4$ GeV. The total Higgs
width $\Gamma_H$ is computed numerically for different values of $m_H$,
with results in agreement with {\tt HDECAY}~\cite{HDECAY1,HDECAY2}.

In \fig{delta vs mH} we plot $\delta$ as a function of $m_H$,
for $\theta = 45^{\circ}$. We see that the interference effect is
stronger for a heavier Higgs boson, and that it reaches $-0.4\%$ 
for $m_H\simeq 150$ GeV.   This mass value is close to the region 
in which there may be sizable contributions to the phase 
from $W$ boson pairs, one on-shell and one off-shell 
in the $H\rightarrow\gamma\gamma$ amplitude;
so the plot cannot be extrapolated much further without performing
this computation.
In general, though, the dominant contribution to $\delta$ for a light
Higgs boson comes from the one-loop $\gamma\gamma\rightarrow b\bar{b}$ 
and $H\rightarrow b\bar{b}$ amplitudes.

\begin{figure}[t]
\begin{center}
\rotatebox{270}{\scalebox{0.63}{
\includegraphics{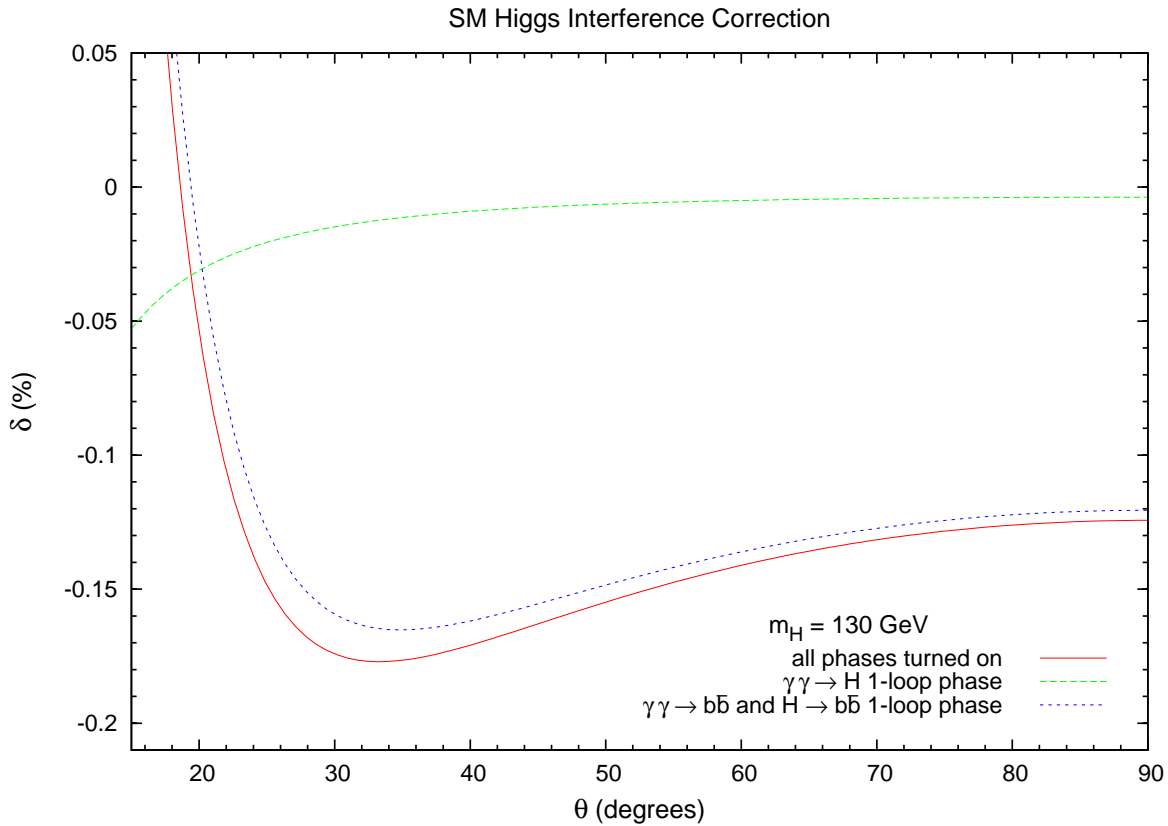}}}
\caption{{\small The percentage reduction of the SM Higgs signal as
a function of the scattering angle for $m_H = 130$ GeV. The solid
curve represents the result with all phases turned on; the dashed
curves turn on different component phases each
time. The total effect is maximized close to $\theta\simeq
35^{\circ}$.}} \label{delta vs theta}
\end{center}
\end{figure}

In \fig{delta vs theta} we plot $\delta$ as a function of the
scattering angle $\theta$, for $m_H = 130$ GeV.  Note that the
small-mass approximation formula~(\ref{delta approx}) for
$\delta$ diverges for small angles.  This behavior can be understood
as coming from the $\gamma\gamma\rightarrow b\bar{b}$ continuum
amplitude, which exhibits a similar angular dependence. 
Keeping the exact $b$-quark mass dependence, using \eqn{ImExactMass},
the divergence is regulated.
We find that for $m_H = 130$ GeV, $\delta = 18\%$ at $\theta = 3^\circ$, 
and that it rolls off to a constant $\delta \approx 35\%$ for 
$\theta < 0.5^\circ$.
Of course it would be very challenging experimentally 
to search for $b$ jets in this far-forward region, and the reason $\delta$ is
increasing is because the continuum $b\bar{b}$ background is increasing.
Away from the forward region, the interference effect has the opposite
sign, negative, and its magnitude becomes maximum for
$\theta\simeq 35^{\circ}$, with $\delta\simeq -0.18\%$.
Again, the phase arising from the one-loop 
$\gamma\gamma\rightarrow b\bar{b}$ and $H\rightarrow b\bar{b}$ 
amplitudes almost solely determines the size of the correction.

In models beyond the SM, such as the MSSM, the coupling of a Higgs
boson to $b$ quarks and to photons is modified.  How does the
interference effect depend on these couplings?   Looking at
\eqn{delta approx}, we see that the two powers of the Yukawa
coupling $\lambda_b\equiv m_b/v$ from
$\left|\mathcal{A}^\mathrm{tree}_{H\rightarrow b\bar{b}}\right|^2$
cancel against the ones contained in $\Gamma_H$ (which for most of the
relevant range of $m_H$ is dominated by the $H\rightarrow b\bar{b}$
decay). There is one extra power of $\lambda_b = m_b/v$ coming from the
$H\rightarrow b\bar{b}$ amplitude in the numerator in \eqn{delta simplified}, 
so the dominant contribution to $\delta$ is linear in $\lambda_b$. 
The subdominant contribution from 
$\mathrm{Im}\{ \mathcal{A}^{(1)}_{\gamma\gamma\rightarrow H} \}$ includes
one more factor of $\lambda_b$, so it is quadratic in $\lambda_b$.

At a photon collider, the unperturbed peak height is proportional 
to the product
$\Gamma(H\rightarrow\gamma\gamma)\times\mathrm{Br}(H\rightarrow b\bar{b})$.  
The $H\rightarrow\gamma\gamma$ width does not depend strongly on
$\lambda_b$ until it gets very large. 
The $H\rightarrow b\bar{b}$ branching ratio is $\approx 1$, 
getting even closer to 1 as $\lambda_b$ increases.
Thus the unperturbed peak height does not change dramatically,
but the fractional shift $\delta$ can increase considerably 
as $\lambda_b$ grows.  In particular for the MSSM, the Yukawa
coupling to the lightest Higgs $h$ is 
$(m_b/v) \times (\sin\alpha/\cos\beta)$,
where $\alpha$ is a Higgs mixing angle and the ratio of vacuum expectation
values of $H_u$ and $H_d$ is $\tan\beta$.
If the heavier Higgs bosons are not decoupled, and $\tan\beta$ is large
(perhaps as large as $\sim$\,50), as in the so-called 
``intense coupling regime''~\cite{BoosIntense1,BoosIntense2},
then $\delta$ can receive a big enhancement.
As an example, we have computed $\delta$ assuming a factor of 20
increase in $\lambda_b$ over the SM value; we 
obtain $\delta \approx - 4\%$ for $m_H = 130$~GeV and $\theta = 45^\circ$,
with a significant contribution now from 
$\mathrm{Im}\{ \mathcal{A}^{(1)}_{\gamma\gamma\rightarrow H} \}$.
(In the very-strong-coupling regime one might also wish to 
compute corrections to $\delta$ due to phases from rescattering 
via $t$-channel Higgs exchange between the $b$ quarks, 
but we have not done so.)

From \eqn{delta approx}, $\delta$ is inversely proportional to
the $H\gamma\gamma$ coupling, given by \eqn{Hgammagamma}. This
means that an enhancement in $\delta$ could also come from a decrease of
$\mathcal{A}^{\left(1\right)}_{\gamma\gamma\rightarrow H}$, {\it e.g.} 
by opposite-sign contributions from extra particles in the loop. However,
such a decrease will also affect
$\Gamma(H\rightarrow\gamma\gamma)$, and consequently reduce the
total number of events, leading to low statistics in the measurement
of the Higgs partial width in the $\gamma\gamma\rightarrow
H\rightarrow b\bar{b}$ channel.

In conclusion, we have presented results for the 
resonance--continuum interference effect in the $\gamma\gamma\rightarrow
H\rightarrow b\bar{b}$ channel at a photon collider, focusing on a
low-mass ($m_H<160$ GeV) Higgs boson. We obtained our results by computing
the relative phase arising from one-loop QCD corrections, exploiting
the unitarity properties of the corresponding diagrams. We found
that the dominant contribution comes from the one-loop
$\gamma\gamma\rightarrow b\bar{b}$ and $H\rightarrow b\bar{b}$ amplitudes,
and that the
magnitude of the effect in the SM is mostly within the range of 0.1--0.2\%. 
This indicates that such an interference effect is negligible 
for the determination of the properties of the Higgs sector 
in the SM, and probably negligible in most regions of MSSM parameter
space, aside from ``intense coupling'' regions.  
The SM effect is an order of magnitude smaller than the 
experimental precision achievable at a photon collider, and therefore 
poses no worry for the measurement of the Higgs partial width at such a machine.

\begin{center}
\textbf{Acknowledgements}
\end{center}

We would like to thank Michael Peskin for useful discussions,
and Jae Sik Lee, Apostolis Pilaftsis and Michael Spira
for comments on the manuscript.
The Feynman diagrams in the paper were made with
\texttt{JaxoDraw}~\cite{Binosi:2003yf}, based on 
\texttt{AxoDraw}~\cite{Vermaseren:1994je}.  
This research was supported by the US Department of
Energy under contract DE--AC02--76SF00515.

%%%%%%%%%%%%%%%%%%%%%%%%

\end{document}